\documentstyle[12pt]{article}
\begin{document}
\title{Isoscalar dipole mode in relativistic\\
random phase approximation}
\author{D. Vretenar$^{1,2}$, A. Wandelt$^{1}$, and P. Ring$^{1}$
\vspace{0.5 cm}\\
$^{1}$ Physik-Department der Technischen Universit\"at M\"unchen,\\
D-85748 Garching, Germany\\
$^{2}$ Physics Department, Faculty of Science, University of Zagreb\\
10000 Zagreb, Croatia}
\maketitle
\begin{abstract}
The isoscalar giant dipole resonance structure in $^{208}$Pb is 
calculated in the framework of a fully consistent relativistic random 
phase approximation, based on effective mean-field Lagrangians
with nonlinear meson self-interaction terms. The results are 
compared with recent experimental data and with calculations performed
in the Hartree-Fock plus RPA framework. Two basic isoscalar dipole
modes are identified from the analysis of the velocity distributions.
The discrepancy between the calculated strength 
distributions and current experimental data is discussed, as 
well as the implications for the determination of the nuclear 
matter incompressibility.
\end{abstract}
\newpage
The study of the isoscalar giant dipole resonance (IS GDR) might 
provide important information on the nuclear matter compression
modulus $K_{\rm nm}$. This, somewhat elusive, quantity defines
basic properties of nuclei, supernovae explosions, neutron stars
and heavy-ion collisions. The range of values of $K_{\rm nm}$ has 
been deduced from the measured energies of the isoscalar giant
monopole resonance (GMR) in spherical nuclei. The complete 
experimental data set on isoscalar GMR, however, does not limit
the range of $K_{\rm nm}$ to better than $200 - 300$ MeV. Also
microscopic calculations of GMR excitation energies have not
really restricted the range of allowed values for the 
nuclear matter compression modulus. On one hand, modern 
non-relativistic Hartree-Fock plus random phase approximation
(RPA) calculations, using both
Skyrme and Gogny effective interactions, indicate that the value of
$K_{\rm nm}$ should be in the range 210-220 MeV~\cite{Bla.80,BBD.95a}. 
In relativistic mean-field models on the other hand, results of both
time-dependent and constrained calculations suggest that
empirical GMR energies are best reproduced by an effective
force with $K_{\rm nm}\approx 250 - 270$ MeV~\cite{Vre.97,Vre.99,Vre.99a}.

In principle, complementary information about the nuclear 
incompressibility, and therefore by extension about the 
nuclear matter compression modulus, could be obtained from 
the other compression mode: giant isoscalar dipole oscillations. 
In first order the isoscalar dipole mode corresponds to 
spurious center-of-mass motion. The IS GDR is a second order
effect, built on $3\hbar \omega$, or higher configurations.
It can be visualized as a compression wave traveling back 
and forth through the nucleus along a definite direction: 
the "squeezing mode"~\cite{Har.80,Har.81}. There are very
few data on IS GDR in nuclei (the current experimental status 
has been reviewed in Ref.~\cite{Garg.99}). In particular, 
recent results on IS GDR obtained by using inelastic
scattering of $\alpha$ particles have been reported for
$^{208}$Pb~\cite{Dav.97}, and for $^{90}$Zr, $^{116}$Sn, 
$^{144}$Sm, and $^{208}$Pb~\cite{Cla.99}. As in the case
of giant monopole resonances, data on heavy spherical nuclei
are particularly significant for the determination of the 
nuclear matter compression modulus: for example 
$^{208}$Pb. However, recent experimental data on IS GDR
excitation energies in this nucleus disagree: the centroid
energy of the isoscalar dipole strength distribution is
at $22.4 \pm 0.5$ MeV in Ref.~\cite{Dav.97}, while 
the value $19.3 \pm 0.3$ MeV has been reported in 
Ref.~\cite{Cla.99}. In the analysis of 
Ref.~\cite{Dav.97}, the "difference of spectra" technique was
employed to separate the IS GDR from the high-energy
octupole resonance (HEOR) in the $0^{\rm o}
 \rightarrow 2^{\rm o}$ 
$\alpha$-scattering spectrum for $^{208}$Pb. On the other
hand, in the experiment on $^{90}$Zr,
$^{116}$Sn, $^{144}$Sm, and $^{208}$Pb of 
Ref.~\cite{Cla.99}, the mixture of isoscalar $L=1$ (IS GDR) and
$L=3$ (HEOR) multipole strength could not be separated by 
a peak fitting technique. Instead, the data were analyzed
by a multipole analysis of 1 MeV slices of the data over the
giant resonance structure, obtained by removing the underlying
continuum. 

In Ref.~\cite{Garg.99} it has been also pointed out that the 
experimental IS GDR centroid energies, 
and therefore the corresponding values 
of the nuclear incompressibility $K_A$, are not consistent with those
derived from the measured energies of the isoscalar GMR in $^{208}$Pb.
In the sum rule approach to the compression modes~\cite{Stri.82}, 
two different models have been considered
for the description of the collective motion: the hydrodynamical 
model and the generalized scaling model. The assumption of the scaling 
model leads to a difference of more than $40\%$ between the values
of the finite nucleus incompressibility $K_A$, when extracted from the 
experimental energies of the IS GDR and the GMR in $^{208}$Pb.  
A consistent value for $K_A$ can be derived from the experimental
excitation energies, only if the two compression modes are 
described in the hydrodynamical model. The resulting 
value of $K_A \approx 220$ MeV, however, is much too high, and
in fact it corresponds to the nuclear matter compression modulus
$K_{\rm nm}$, as derived from non-relativistic Hartree-Fock plus RPA 
calculations. This is not difficult to understand, since the
expressions for $K_A$ in both models were derived in the limit
of large systems and consequently do not account for surface
effects~\cite{Stri.82}. Both models, however, are approximations
to a full quantum description: the time-dependent Hartree-Fock or,
equivalently, the RPA. Therefore, fully microscopic calculations 
might be necessary, in order to resolve the apparent discrepancy
between the values of $K_A$ extracted from the IS GDR and the 
GMR in $^{208}$Pb. 

Non-relativistic self-consistent Hartree-Fock
plus RPA calculations of dipole compression modes in nuclei 
were reported in the work
of Van Giai and Sagawa~\cite{Giai.81}, and more recently 
in Refs.~\cite{Ham.98} and ~\cite{Colo.99}. A number of 
different Skyrme parameterizations were used in these
calculations, and the result is that all of them systematically
overestimate the experimental values of the IS GDR centroid energies,
not only for $^{208}$Pb, but also for lighter nuclei. In particular,
those interactions that reproduce the experimental excitation 
energies of the GMR (SGII and SKM$^{*}$), predict centroid energies
of the IS GDR in $^{208}$Pb that are $4-5$ MeV higher than those
extracted from small angle $\alpha$-scattering spectra. In 
Ref.~\cite{Colo.99} effects that go beyond the mean-field 
approximation have been considered: the inclusion of the 
continuum and $2p-2h$ coupling. It has been shown that 
the coupling of RPA states to $2p-2h$ configurations,
although it reproduces the total width, results
in a downward shift of the resonance energy of less than 1 MeV
with respect to the RPA value. It appears, therefore, 
that the presently available 
data on excitation energies of the compression modes in nuclei:
the GMR and the IS GDR, cannot be consistently reproduced by 
theoretical models.

In Ref.~\cite{Vre.97} we have performed time-dependent
and constrained relativistic mean-field calculations for 
the monopole giant resonances in a number of spherical
closed shell nuclei, from $^{16}$O to $^{208}$Pb.
It has been shown that, in the framework of
relativistic mean field theory, the
nuclear matter compression modulus $K_{\rm nm} \approx 250
- 270$ MeV is in reasonable agreement with the available
data on spherical nuclei. This value is approximately 20\%
larger than the values deduced from 
non-relativistic density dependent Hartree-Fock
calculations with Skyrme or Gogny forces.
In particular, among the presently available effective 
Lagrangian parameterizations, the NL3 effective force~\cite{LKR.96}
with $K_{\rm nm} = 271.8 $ MeV, provides the best description of 
the mass dependence of the GMR excitation energies. 
Preliminary calculations with the time-dependent
relativistic mean-field model~\cite{VBR.95}, indicate that 
the NL3 effective interaction, which reproduces exactly the 
excitation energy of the GMR in $^{208}$Pb (14.1 MeV), 
overestimates the reported centroid energy of the IS GDR by
at least 4 MeV. However, due to complications arising from
the spurious center-of-mass motion, the time-dependent
relativistic mean-field model computer code develops a
numerical instability which prevents the precise 
determination of the IS GDR excitation energy. In the 
present analysis, therefore, we apply the relativistic
random phase approximation (RRPA) to the description of the
isoscalar dipole oscillations in $^{208}$Pb.

The RRPA represents the small amplitude limit of the
time-dependent relativistic mean-field theory. Self-consistency
will therefore ensure that the same correlations which 
define the ground-state properties, also determine
the behavior of small deviations from the equilibrium.
The same effective Lagrangian generates the Dirac-Hartree
single-particle spectrum and the residual particle-hole 
interaction. Some of the earliest applications of the RRPA
to finite nuclei include the description of low-lying negative 
parity excitations in $^{16}$O~\cite{Fur.85}, and studies
of isoscalar giant resonances in light and medium nuclei~\cite{lhu.89}.
These RRPA calculations, however, were based on the most simple,
linear $\sigma - \omega$ relativistic mean field model. It is well
known that for a quantitative description of ground- and excited 
states in finite nuclei, density dependent interactions have to 
be included in the effective Lagrangian through the meson
non-linear self interaction terms. The RRPA response functions with
nonlinear meson terms have been derived in Refs.~\cite{ma.97,ma.97a},
and applied in studies of isoscalar and isovector giant resonances.
However, the calculated excitation energies did not reproduce
the values obtained with the time-dependent relativistic mean-field
model~\cite{Vre.97,VBR.95}. The reason was that the RRPA configuration
spaces used in Refs.~\cite{ma.97,ma.97a} did not include the 
negative energy Dirac states. In Ref.~\cite{Daw.90} it has been
shown that an RRPA calculation, consistent with the mean-field
model in the $no-sea$ approximation, necessitates configuration 
spaces that include both particle-hole pairs and pairs formed
from occupied states and negative-energy states. The contributions
from configurations built from occupied positive-energy states and
negative-energy states are essential for current conservation and 
the decoupling of the spurious state. In addition, configurations
which include negative-energy states give an important contribution
to the collectivity of excited states. In a recent study~\cite{Wa.00}
we have shown that, in order to reproduce results of time-dependent
relativistic mean-field calculations for giant resonances, the 
RRPA configuration space must contain negative-energy Dirac states, 
and the two-body matrix elements must include contributions
from the spatial components of the vector meson fields.
The effects of the Dirac sea on the excitation energy 
of the giant monopole states have been also recently studied in an 
analytic way within the $\sigma - \omega$ model~\cite{KS.00}.

In Fig.~1 we display the IS GDR strength distributions in $^{208}$Pb:
\begin{equation}
B^{T=0}(E1, 1_i \rightarrow 0_f)  =  \frac{1}{3} \, |
\langle 0_f || \hat{Q}^{T=0}_1 || 1_i \rangle |^2,
\end{equation}
where the isocalar dipole operator is
\begin{equation}
\hat{Q}^{T=0}_{1 \mu}  =  e \, \sum^{A}_{i=1} \,
\gamma_0 \, (r^3 - \eta r) \ Y_{1 \mu}(\theta_i, \varphi_i),
\label{operator}
\end{equation}
and
\begin{equation}
\eta  =  \frac{5}{3} \, < r^2 >_{_0} \quad .
\label{eta}
\end{equation}
The calculations have been performed within the framework of the
self-consistent Dirac-Hartree plus relativistic RPA. The effective
mean-field Lagrangian contains nonlinear meson self-interaction terms,
and the configuration space includes both particle-hole pairs, and pairs 
formed from hole states and negative-energy states. The choice of 
the dipole operator (\ref{operator}), with the parameter $\eta$ 
determined by the condition of translational invariance, ensures that
the IS GDR strength distribution does not contain spurious 
components that correspond to the center-of-mass motion~\cite{Giai.81}.
The strength distributions in Fig.~1 have been calculated with the
NL1 ($K_{\rm nm} = 211.7$ MeV)~\cite{RRM.86}, 
NL3~\cite{LKR.96} ($K_{\rm nm} = 271.8$ MeV), 
and NL-SH ($K_{\rm nm} = 355.0$ MeV)~\cite{SNR.93}
effective interactions. These three forces, in order of increasing
values of the nuclear matter compressibility modulus, have been 
extensively used in the description of a variety of properties
of finite nuclei, not only those along the valley of $\beta$-stability,
but also of exotic nuclei close to the particle drip lines. 
In particular, in Ref.~\cite{Vre.97} it has been shown
that the NL3 ($K_{\rm nm} = 271.8$ MeV) effective interaction 
provides the best description of experimental data on 
isoscalar giant monopole resonances.

The calculated strength distributions are similar to those 
obtained within the non-relativistic Hartree-Fock plus RPA
framework, using Skyrme effective forces~\cite{Giai.81,Colo.99}.
In disagreement with reported experimental results, all theoretical
models predict a substantial amount of isoscalar dipole
strength in the $8 - 14$ MeV region. The centroid energies 
of the distributions in the high-energy region between 20 and
30 MeV, are $4 - 5$ MeV higher than those extracted
from the experimental spectra. It also appears that the
centroid energies of the low-energy distribution do not 
depend on the nuclear matter incompressibility of the 
effective interactions. On the other hand, 
the IS GDR strength distributions 
in the low-energy region display the expected
mass dependence. We have also performed calculations for 
a number of lighter spherical nuclei, and verified 
that with increasing mass the centroid is 
indeed shifted to lower energy. 
When comparing with experimental data, it should be
pointed out that the usable excitation energy
bite in the experiment reported in Ref.~\cite{Dav.97} was
$14 - 29$ MeV, and therefore a low-energy isoscalar dipole
strength could not be observed. In this respect, somewhat
more useful are the data from the experiment 
reported in Ref.~\cite{Cla.99}, where spectra in the 
energy range $\approx 4 < E_x < 60$ MeV have been observed.
The results of an DWBA analysis of the experimental spectra,
however, attribute the isoscalar strength in the $10 - 15$ MeV 
region exclusively to the giant monopole (GMR) and giant 
quadrupole (GQR) resonances. It should be emphasized that
a possible excitation of isoscalar dipole strength in this
energy region and its interference with the GQR
cannot be excluded~\cite{Har.80}.

In the high energy region the calculated dipole strength 
exhibits the expected dependence on the nuclear matter
compressibility modulus of the effective interactions 
(NL1, NL3, NL-SH). The centroid of the strength distribution
is shifted to higher energy with increasing values of 
$K_{\rm nm}$. These energies, however, are considerably 
higher than the corresponding experimental IS GDR 
centroids~\cite{Dav.97,Cla.99}. Though, in order to 
precisely determine the IS GDR excitation energy from the 
experimental spectrum, the dipole strength has to be 
separated from the high-energy octupole resonance (HEOR),
and this is not always possible~\cite{Cla.99}. Using 
the NL3 effective interaction, we have calculated the 
octupole strength distribution. The centroid of the 
HEOR is found at $\approx 22$ MeV, well below the 
IS GDR main peak, but more than 2 MeV above the
experimental value for the HEOR centroid~\cite{Dav.97}.
Incidentally, our calculated HEOR peak approximately coincides
with the experimental value of the IS GDR centroid~\cite{Dav.97}. 

The IS GDR transition densities for $^{208}$Pb
are shown in Fig.~2. The transition densities correspond
to the NL3 strength distribution in Fig.~1. Since it appears
that none of the effective interactions reproduces the 
experimental position of the IS GDR, the remainder of the 
present analysis will be only qualitative, and we choose to
display only results obtained with the NL3 set of Lagrangian
parameters. On the qualitative level, the other two
effective interactions produce similar results.
In Fig.~2. we plot proton (dot-dashed), neutron (dashed), 
and total (solid) transition densities for
two representative peaks from Fig.~1:  
10.35 MeV (a) is the central peak in the low-energy
region, and 26.01 MeV (b) is the energy of the main peak
in the region above 20 MeV. The transition densities for 
both peaks exhibit a radial dependence characteristic 
for the isoscalar dipole mode, and they can be compared
with the corresponding transition densities in the 
scaling model, or with those which result from constrained 
calculations~\cite{Stri.82}. While for the high-energy peak 
the proton and neutron transition densities display 
an almost identical radial dependence, the pattern is 
more complicated for the peak at 10.35 MeV.

RPA calculations, therefore, predict the fragmentation of the 
isoscalar dipole strength distribution into two broad structures:
one in the energy window between $8 - 14$ MeV, and the
other in the high-energy region around $\approx 25$ MeV.
The position of the low-energy structure does not depend
on the compressibility modulus, i.e. it does not correspond 
to a compression mode. Additional information on the underlying
collective dynamics can be obtained through a study of 
transition currents. In Fig.~3 we plot the velocity fields
for the two peaks at 10.35 MeV (a) and 26.01 MeV (b).
The velocity distributions are derived from the corresponding
transition densities, following the procedure described in 
Ref.~\cite{Serr.83}. The "squeezing" compression mode is 
identified from the flow pattern which corresponds to 
the high-energy peak at 26.01 MeV. The flow lines 
concentrate in the two "poles" on the symmetry axis at
$z \approx \pm 2.5$ fm. The velocity field corresponds 
to a density distribution which is being compressed in the
lower half plane, and expands in the 
upper half plane. The centers of compression and expansion
are located on the symmetry axis, at approximately half 
the distance between the center and the surface of the nucleus.
It is obvious that the excitation energy of this mode will
strongly depend on the compressibility modulus. The flow 
pattern for the lower peak at 10.35 MeV is very different. 
The flow lines describe a kind of toroidal motion, which 
is caused by the surface effect of the finite nucleus.
The density wave travels through the nucleus along the
symmetry axis. The reflection of the wave on the surface, 
however, induces radial components in the velocity field.
Although it corresponds to dipole oscillations, this is
not a compression mode. We have verified that also other
dipole states in this energy region display similar 
velocity fields.

In conclusion, the isoscalar giant dipole resonance  
in $^{208}$Pb has been 
calculated in the framework of the relativistic RPA,
based on effective mean-field Lagrangians
with meson self-interaction terms. The results
have been compared with recent experimental data and with
calculations performed in the Hartree-Fock plus RPA framework.
While the results of the present RRPA study are consistent
with previous theoretical analyses, they strongly disagree
with reported experimental data on the position of the 
IS GDR centroid energy in $^{208}$Pb. This is a serious
problem, not only because the disagreement between theory 
and experiment is an order of magnitude larger than 
for other giant resonances, but also because the present
data on IS GDR are not consistent with the value
of the nuclear incompressibility $K_A$ derived from
the measured excitation energy of the isoscalar GMR.
This inconsistency could, perhaps, be explained by 
a possible excitation of isoscalar dipole strength in the
low-energy window between 8 MeV and 14 MeV. Although
predicted by all theoretical models, the low-lying 
IS GDR strength is not observed in the experimental spectra.
From the analysis of the velocity fields, we have identified
two basic isoscalar dipole modes. The "squeezing" compression 
mode is found in the high-energy region at $\approx 26$ MeV.
The low-energy dipole mode does not correspond to a 
compression mode, and its dynamics is determined by 
surface effects.
\bigskip\\
\noindent
{\bf Acknowledgments}

We thank P.F. Bortignon, G.Col\`o, U. Garg, Z.Y. Ma, 
and N. Van Giai for useful comments.
This work has been supported in part by the
Bundesministerium f\"ur Bildung und Forschung under
contract 06~TM~875.

\newpage
{\bf Figure Captions}

\begin{itemize} 
\item{\bf Fig.1} IS GDR strength distributions in $^{208}$Pb calculated
with the NL1 (dashed), NL3 (solid), and NL-SH (dot-dashed) effective 
interactions.

\item{\bf Fig. 2} IS GDR transition densities for $^{208}$Pb calculated
with the NL3 parameter set. Proton (dot-dashed), neutron (dashed), and
total (solid) transition densities are displayed for the peaks at
10.35 MeV (a) and 26.01 MeV (b).

\item{\bf Fig. 3} Velocity distributions for the two isoscalar dipole 
modes in $^{208}$Pb calculated with the NL3 effective interaction.
The velocity fields correspond to the two peaks at
10.35 MeV (a) and 26.01 MeV (b).
\end{itemize}
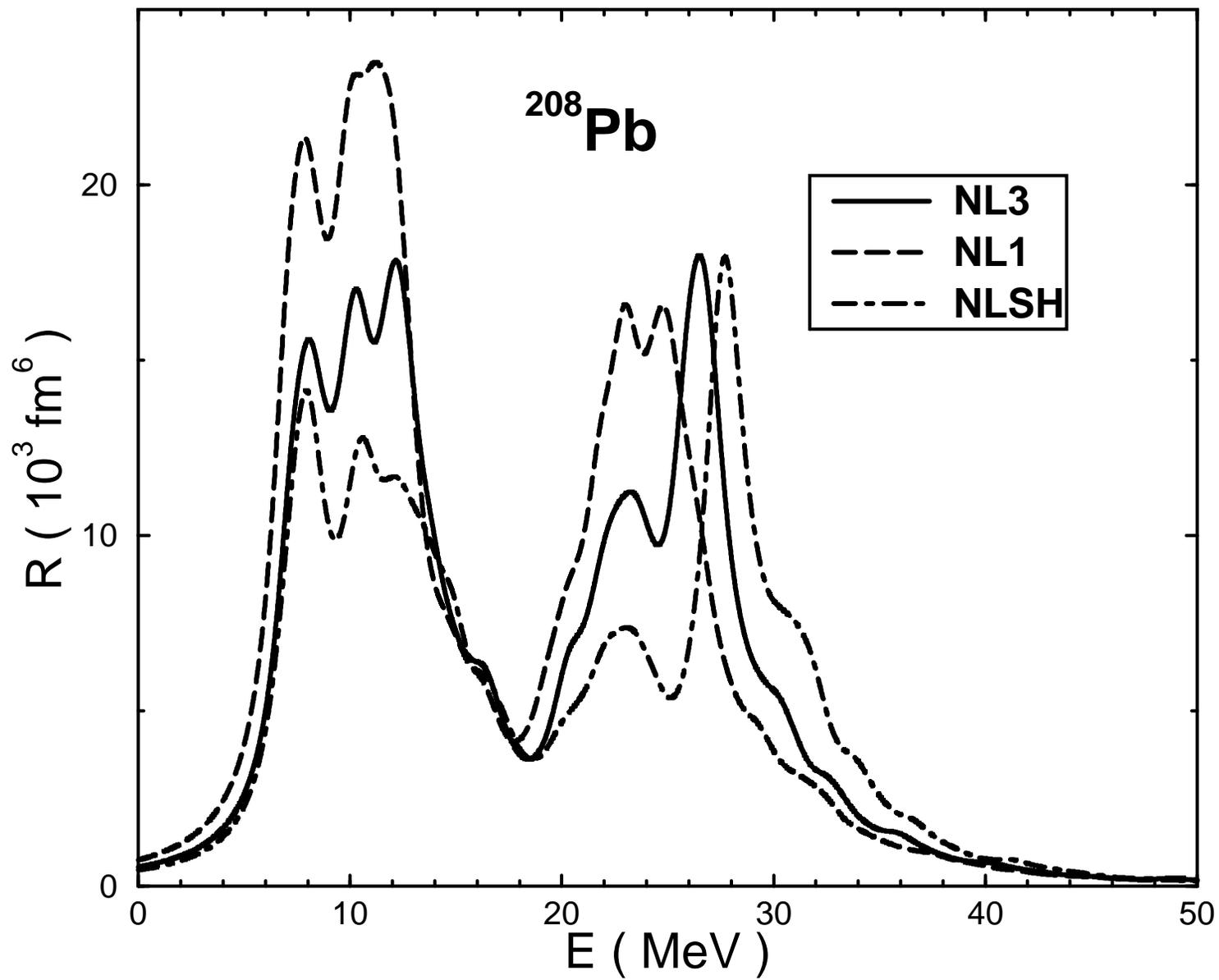
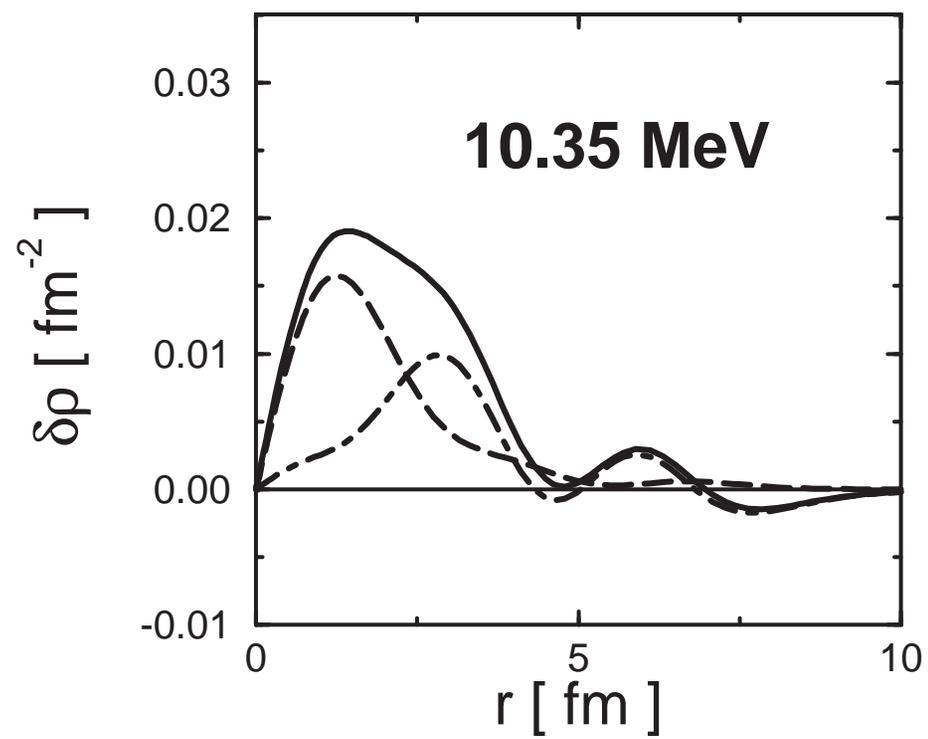
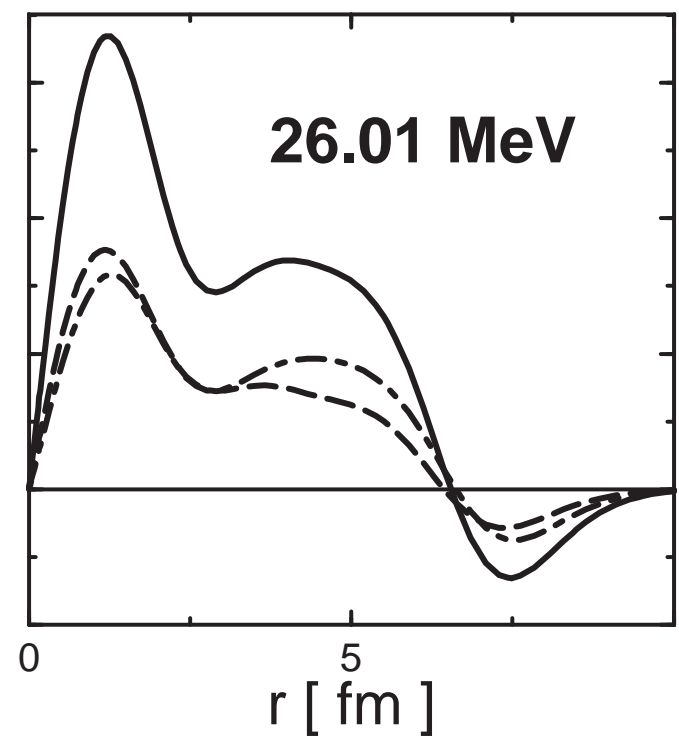
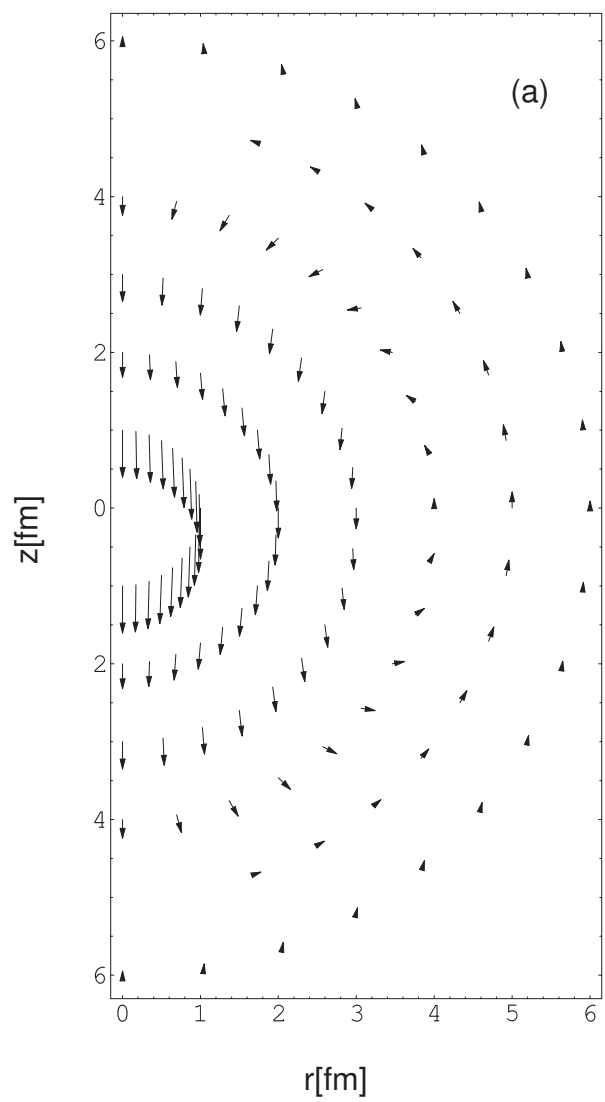
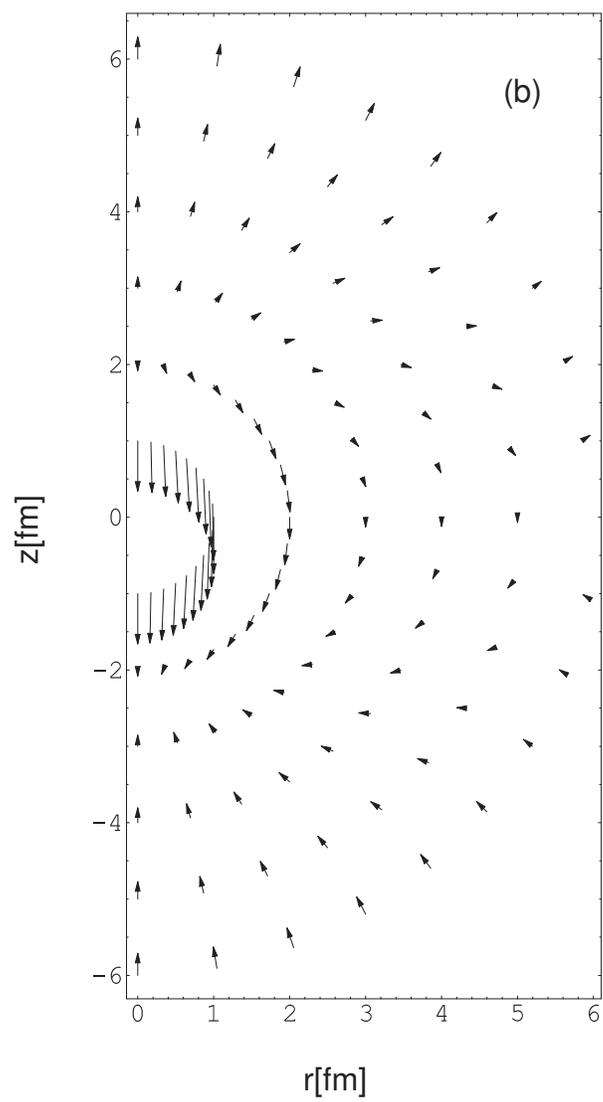
\end{document}